\documentclass[nofootinbib,twocolumn]{revtex4}
\usepackage{graphicx}
\usepackage{amssymb}
\usepackage{amsmath}
\usepackage{bm}
\usepackage{hyperref}
\usepackage{epstopdf}
\begin{document}

\title{Non-linear clustering during the BEC dark matter phase transition}

\author{Rodolfo~C.~de~Freitas}\email{rodolfo.camargo@pq.cnpq.br}
\affiliation{Universidade Federal do Esp\'irito Santo, Av. Fernando Ferrari, Goiabeiras, Vit\'oria, Brasil}
\affiliation{Instituto Federal de Educa\c{c}\~ao, Ci\^encia e Tecnologia do Esp\'irito Santo, Avenida Vit\'oria 1729, Jucutuquara, Vit\'oria, Brazil}
\author{Hermano~Velten}\email{velten@pq.cnpq.br}
\affiliation{Universidade Federal do Esp\'irito Santo, Av. Fernando Ferrari, Goiabeiras, Vit\'oria, Brasil}
\affiliation{CPT, Aix Marseille Universit\'e, UMR 7332, 13288 Marseille,  France}

\begin{abstract} 
Spherical collapse of the Bose-Einstein Condensate (BEC) dark matter model is studied in the Thomas Fermi approximation. The evolution of the overdensity of the collapsed region and its expansion rate are calculated for two scenarios. We consider the case of a sharp phase transition (which happens when the critical temperature is reached) from the normal dark matter state to the condensate one and the case of a smooth first order phase transition where there is a continuous conversion of ``normal'' dark matter to the BEC phase. We present numerical results for the physics of the collapse for a wide range of the model's space parameter, i.e. the mass of the scalar particle $m_{\chi}$ and the scattering length $l_s$. We show the dependence of the transition redshift on $m_{\chi}$ and $l_s$. Since small scales collapse earlier and eventually before the BEC phase transition the evolution of collapsing halos in this limit is indeed the same in both the CDM and the BEC models. Differences are expected to appear only on the largest astrophysical scales. However, we argue that the BEC model is almost indistinguishable from the usual dark matter scenario concerning the evolution of nonlinear perturbations above typical clusters scales, i.e., $\gtrsim 10^{14}M_{\odot}$. This provides an analytical confirmation for recent results from cosmological numerical simulations [H.-Y. Schive {\it et al.}, Nature Physics, {\bf10}, 496 (2014)].  \\


\end{abstract}

\maketitle

\section{Introduction}

It is widely accepted that dark matter is one of the main components of the universe. Due to the strong observational evidence corroborating its existence many different areas of physics have incorporated dark matter related investigations in their agenda. According to the standard cosmological model, dark matter composes around $1/4$ of the universe's energy budget and $5/6$ of the total matter. Baryons represent the remaining fraction of the latter. This picture has been confirmed by different data, but remarkably by the latest {\it Planck} results \cite{PlanckCosmoParam}.

The crucial aspects of these studies concern the particle nature and the astrophysical/cosmological behavior of such component. At the particle level, candidates belonging to the WIMP (weakly interacting massive particles) category produce a viable model (see \cite{Baer}---and references therein---for a very recent review). Also, for the homogeneous, isotropic and expanding background, the dark matter ensemble should present a vanishing pressure in order to enable structure formation \cite{cdmpressure}.

Although the success of the Cold Dark Matter (CDM) scenario it is important to mention some of its drawbacks. The theoretical clustering patterns (calculated via numerical simulations) of CDM particles at galactic level correspond to the NFW profile \cite{NFW} which is cuspy at the centre of the particle distribution. This seems to be in clear contradiction to the observed velocities in the central region of galaxies which demand a cored distribution. At the same time, the simulated distribution of satellites around typical Milk Way like galaxies shows one order of magnitude excess of sub-structures which are not observed. These two issues are known as the cusp-core problem and the missing satellites problem, respectively. Even if baryonic physics in such simulations could eventually alleviate these problems, it is not clear so far whether or not CDM is the correct model for the dark matter phenomena. See \cite{gov} and references therein. 

One can argue that dark matter is a pathological manifestation of choosing Einstein's general relativity (GR) as the gravitational theory. This suspect is the pillar of a research line in which modified gravity theories are invoked. See \cite{Mod} for reviews on modified gravity models and their observational constraints. However, reliable experiments at the solar system level confirm GR predictions with great accuracy \cite{Will}. Therefore, this fact seems to be powerful enough to keep in a fist moment GR as our standard description for gravitational interaction.

Since there are no confirmed evidence to abandon GR, dark matter remains being essential and therefore one needs new alternatives within this context. In this case, the possibilities are also vast. The classical ones were hot dark matter (HDM) \cite{HDM} and warm dark matter (WDM) \cite{wdm}. While the former has been ruled out due to the positive observation of galaxies below the jeans mass scale of relativistic dark matter particles, the latter is one of the leading rivals of CDM. Indeed, particles with masses $m\sim$ keV fit the WDM spirit. They are not as light as HDM particles and therefore allowing the existence of structures and, at the same time, not as heavy as CDM, in such a way that there would exist some suppression mechanism able to alleviate the small scale problems of the CDM paradigm (see however \cite{wdm2} for a recent discussion of WDM results). Models with a similar clustering dynamics as WDM are, for instance, fuzzy dark matter \cite{fuzzy}, the self interacting dark matter \cite{sidm} and the viscous dark matter \cite{vdm}.      

In this work we study a dark matter model which has a different nature. Let us assume $0-$spin DM particles having therefore a bosonic distribution. As predicted and already observed in laboratory bosonic particles are able to condensate \cite{BEClab} (see also \cite{BEClab2}), occupying the same energy state and forming the so-called Bose-Einstein condensates when their temperature reaches the critical value $T_{crt}$. Of course, this phenomenon occurs under very controlled experimental situations, but one might wonder in principle what happens if the same would happen on astrophysical scales. 

Although quite hypothetical this description could serve as an effective approach for understanding dark matter as a cosmological scalar field $\phi$ whose dynamics is driven by some repulsive potential $V(\phi)$. This gives rise to the Bose-Einstein Condensate (BEC) dark matter model which has been widely studied \cite{Siki, Bohmer, harko1, Abril, Maxim, LiShapiro}. The main idea is that normal, i.e., non-condensate, dark matter undergoes a phase transition at some critical redshift $z_{crt}$ during the universe's evolution. Then, independently on the details of the transition, all the dark matter converts into the condensate state forming a BEC ``fluid'' \footnote{A recent controversial claim challenging the existence of such astronomical BEC condensates has been discussed in \cite{Guth}.}. 

The dynamics of BEC systems is studied via the Gross-Pitaevskii equation, which is a nonlinear Schrodinger equation \cite{BEC}. From this starting point, the Madelung decomposition is used to transform the BEC dynamics into a set of fluid equations resulting in an effective positive pressure. With such fluid picture one is able to investigate astrophysical/cosmological problems. This procedure will be shown in more details in the next section.

The general aspects of this model concerning the background evolution and the linear perturbations are already very well understood \cite{MaximZel, AbrilSF, wamba, Chavanis, BenKain, rodolfo, Alcu}. But, in order to fully understand the final clustering patterns of the BEC dark matter model high resolution hydrodynamical/N-body simulations are still needed \cite{simu}. More recently, Ref. \cite{Mocz} has formulated smoothed-particle hydrodynamics numerical methods to solving general Gross-Pitaevskii-Poisson system. Schive et al \cite{Schive} provided recently high-resolution cosmological simulations for the model. They obtained that there is a remarkable difference at the internal galactic level, i.e., its density profile. The latter result is indeed desired. However, they found that BEC DM is indistinguishable from CDM at large cosmological scales. Our focus here in this work is to understand such latter claim. From the theoretical point of view, a first step on this issue is the study of the nonlinear gravitational collapse in a cosmological background. Concerning the BEC dark matter model, recently Ref. \cite{HarkoBECCollapse} addressed the collapse of ``already formed BEC condensates'', i.e., only the post-transition stage. Nevertheless, a realistic configuration can be more complicated since it also involves the dynamics of the baryonic component as the universe evolves from the matter to the dark energy domination epochs. Moreover, the phase transition can also take place during the evolution of the collapsed region. Therefore, especially for galaxy cluster scales, the evolution of the background cosmological dynamics should be taken into account. 

We will perform in this work a natural extension of Ref. \cite{HarkoBECCollapse} which has analysed the ``free-fall'' collapsed of a BEC dark matter sphere. However, we assume a more realistic cosmological scenario where dark matter coexist with baryons and a cosmological constant. Then, we address the correct case where the transition occurs during the nonlinear clustering process. 

Fundamental quantities here are the condensate parameters, namely, the mass of the particle $m_{\chi}$ and the scattering length $l_s$. They determine the moment at which the phase transition takes place $z_{crt}$ and the speed of sound in the condensate fluid, for example. After the critical redshift $z_{crt}$ one can admit two different dynamics.
The simplest case is to assume an abrupt transition, i.e., for $z<z_{crt}$ all dark matter obeys the Bose-Einstein dynamics. This seems to be a reasonable approximation to the problem. This situation will be studied in section \ref{SectionIII}. 

One can also assume the case in which the full conversion of all dark matter occurs in a finite time and it finishes at a redshift $z_{BEC}<z_{crt}$. Therefore, the phase transition lasts a finite time in which a mixture of ``normal'' and condensate dark matter make up the total matter component. We study in section \ref{SectionIV} this case.

We present our results covering many order of magnitude in the model parameter space $10^{-6}$ meV$< m_{\chi} < 10^{4}$ meV$ ; 10^{-12}$ fm$ < l_s < 10^{12} $ fm. Interesting quantities to be found here are the final (at $z=0$) value of the density contrast and the expansion rate and the redshift of the turnaround $z_a$, i.e., the moment at which the collapsed region detaches from the background. 

In summary, this paper has the following structure. In the next section we develop the background dynamics of the BEC dark matter. We present in section \ref{SectionIII} general equations for the spherical top-hat collapse formalism. These equations will be studied in more detail in sections \ref{SectionIV} and \ref{SectionV} where, respectively, we address the case of abrupt transition and the usual phase transition. We conclude in the final section.

\section{The background dynamics of the Bose-Einstein Condensate dark matter}

In this work we always have a flat background dynamics composed by baryons, dark matter and a cosmological constant. This expansion rate reads 
\begin{equation}
H^2=\frac{8\pi G}{3}\left(\rho_b+\rho_{dm}+\rho_{\Lambda}\right).
\end{equation}
The post-decoupling dynamics of the baryonic component is assumed to be pressureless $P_b=0$ and therefore $\rho_b=\rho_{b0}(1+z)^3$ where $\rho_{b0}$ is its density today at $z=0$. Its value is such that $\rho_{b0}=\Omega_{b0} \rho_{c0}$, where $\rho_{c0}=3H^2_0 / 8\pi G$. We can safely adopt $\Omega_{b0}=0.05$ according to nucleosynthesis constraints. The Hubble constant assumed here is $H_0=70$ Km/s/Mpc. We will also fix $\Omega_{dm0}=0.25$ or equivalently $\Omega_{\Lambda}=0.75$. 
  
The difference here from the standard $\Lambda$CDM model will be the dark matter dynamics. Before the transition takes place, at temperatures $T>T_{crt}$; or redshifts $z>z_{crt}$, DM behaves as an isotropic gas in thermal equilibrium. From kinetic theory the pressure of a non-relativistic gas in this regime is given by
\begin{equation}\label{pdm}
p_{dm}=\frac{g_s}{3h^3}\int\frac{q^2 c^2}{E}f(q)d^3p\approx4\pi\frac{g_s}{3h^3}\int\frac{q^4}{m}\rightarrow\sigma^2 \rho_{dm},
\end{equation}
with $\sigma^2=\left\langle v^2\right\rangle /3 c^2$, where $g_s$ is the number of spin degrees of freedom, $h$ the Planck constant, $q$ the momentum of a particle with energy $E=\sqrt{q^2c^2+m^2c^4}$ and distribution function $f$. A typical value for the velocity dispersion is $\sigma=3 \times 10^{-6}$. In practice, since this quantity can be seen as the dark matter equation of state parameter $w_{dm}=p_{dm}/\rho_{dm}$ this value is consistent with the assumption of pressureless fluid usually adopted for CDM. Note that the full relativistic fluid is obtained when $\left\langle v^2\right\rangle = c^2$.

After dark matter's conversion it obeys the condensate dynamics, which is governed by the Gross-Pitaevskii equation
\begin{equation}\label{GP}
i \hbar \frac{\partial \Psi}{\partial t}=-\frac{\hbar^2}{2 m_{\chi}}\nabla^2 \Psi + V(r,t)\Psi +g(\left|\Psi\right|) \Psi,
\end{equation}
where $m_{\chi}$ is the mass of the particle and $V(r,t)$ is the trapping potential. The non-linearity term with only two-body interparticle interaction (quadratic) reads
\begin{equation}
g(\left|\Psi\right|)=U_0 \left|\Psi\right|^2,
\end{equation}
where $U_0=4\pi \hbar^2 l_s/m^3_{\chi}$. This definition has the fundamental parameters of the model, namely the scattering length $l_s$ and the particle mass $m_{\chi}$. The former is associated to the nature of the short range self-interactions in the condensate. For example, in laboratory systems, it can be either positive (the case of Rb$^{87}$ atoms with $l_s=5.45$ nm and then repulsive interactions) \cite{Ju} or negative (the case of Li$^7$ atoms with $l_s=−1.45$ nm and then attractive interactions) \cite{Abdu}. In this work we will consider only cases where $l_s>0$. The impact of $l_s$ on the mass-radius configurations of astrophysical BEC has been investigated in \cite{delfini}.

Note that there appears some degeneracy for the $U_0$ parameter, i.e., there are infinities combinations of $l_s$ and $m$ capable to produce the same $U_0$ value. We discuss this degeneracy and the admissible numerical values of these parameters in the next sections.

In order to apply the Gross-Pitaevskii equation to astrophysical problems one proceeds with the so-called Madelung decomposition. In this procedure, the wave function is replace by
\begin{equation}
\Psi=\sqrt{\rho(r,t)}\, e^{\frac{i}{\hbar}S(r,t)},
\end{equation}
where $\rho=\left|\Psi\right|^2$ is the number density of the system and $S$ is the velocity potential. The mass/energy density can be written in terms of the mass of each individual particle as $\rho_{\chi}=m_{\chi}\rho$. 

Therefore, the BEC system can be described in terms of a hydrodynamical set of equations, which are
\begin{eqnarray}
   & & \frac{\partial\vec{u}}{\partial t}+\left(\vec{u}\cdot\nabla\right)\vec{u}=-\frac{\nabla p_{\chi}}{\rho_\chi}-\nabla\left(\frac{V}{m_\chi}\right)-\frac{\nabla Q}{m} \,, \\
	 & & \frac{\partial \rho_\chi}{\partial t}+\nabla\cdot\left(\rho_{\chi} \vec{u}\right)=0 \,,
\end{eqnarray}
where we define
\begin{eqnarray}
   \vec{u} &=& \frac{\hbar}{m_\chi}\nabla S \,, \\
	 Q &=& -\frac{\hbar^2}{2m_{\chi}}\frac{\nabla^2\sqrt{\rho_\chi}}{\sqrt{\rho_\chi}} \,.
\end{eqnarray}

The particle self-interaction of this specific BEC-inspired fluid gives rise to a pressure of polytropic form
\begin{equation}\label{Pbec}
p_{\chi}=\frac{2\pi\hbar^2 l_s}{m^3_{\chi}}\rho^2_{\chi}.
\end{equation}
On the other hand, the quantum potential $Q/m_\chi$ results in the often called quantum pressure\footnote{Note that both quantum pressure and self-interaction pressure are of quantum mechanical origin.}. We can use the identity
\begin{equation}
   \frac{\partial_{j}p_{ij}}{\rho_\chi} \equiv \frac{\partial_i Q}{m_\chi} \,,
\end{equation}
where $p_{ij}$ is the quantum anisotropic pressure tensor \cite{delfini}, given by
\begin{equation}
   \label{anisotropic}
   p_{ij} = \frac{\hbar^2}{2m^{2}_{\chi}}\left(\frac{\partial_i\rho\partial_j\rho}{\rho_\chi}-\delta_{ij}\nabla^2\rho\right) \,.
\end{equation}

For the problem we have in mind, the potential $V(r,t)$ in (\ref{GP}) is in fact the gravitational potential which is sourced by $\rho_{\chi}$ via the Poisson equation $\nabla^2 V = 4 \pi G \rho_{\chi}$. This allows us to solve the system of equations.

In the cases where the pressure due to the particle self-interaction dominates, the quantum anisotropic pressure can be neglected. This is the so called Thomas-Fermi approximation. In \cite{RindlerDaller} the authors estimated, for the case of BEC dark matter halos, in which cases the Thomas-Fermi limit is valid. They consider the forces associated with both pressures, that balances the gravitational collapse, and find that the Thomas-Fermi regime is valid when
\begin{equation}
   \frac{\kappa}{\kappa_H} \gg 2 \, ,
\end{equation} 
where
\begin{equation}
   \kappa = 4\pi \hbar^2\frac{l_s}{m_\chi} \,. \\
\end{equation}
Adopting $R$ as the mean radius and $M$ as the mass of a BEC dark matter halo it is found that

\begin{equation}
\label{kappaH}
\kappa_H = \frac{2}{3}\pi \hbar^2 \frac{R}{M} \,.
\end{equation}

The quantity (\ref{kappaH}) can be written in characteristic values
\begin{equation}
   \kappa_H = 2.252\times 10^{-64}\left(\frac{R}{100~\textup{kpc}}\right)\left(\frac{10^{12}~M_{\odot}}{M}\right)~\textup{eV}~\textup{cm}^3 \,.
\end{equation}
If we consider halos with a size between the Milky Way ($M=10^{12}~M_{\odot}$ and $R=100~\textup{kpc}$) and a typical dwarf galaxy ($M=10^{10}~M_{\odot}$ and $R=10~\textup{kpc}$) we can constraint $\kappa_H$ in the range
\begin{equation}
   \kappa_H \approx 2\times \left(10^{-64} \textrm{ -- } 10^{-63}\right) ~\textup{eV}~\textup{cm}^3 \,.
\end{equation}
Using the model parameters range which will be adopted in this work ($10^{-6}$ meV$< m_{\chi} < 10^{4}$ meV$ ; 10^{-12}$ fm$ < l_s < 10^{12} $ fm) we calculate that
\begin{equation}
   \kappa \approx 2\times \left(10^{-43} \textrm{ -- } 10^{27}\right) ~\textup{eV}~\textup{cm}^3 \,,
\end{equation}
which indicates that the Thomas-Fermi approximation can be adopted.

Another comment about the justification of the use of the Thomas-Fermi approximation relies on the fact that we are going to focus on the largest cosmological scales. For example, in the Fourier space density perturbations are affected by the quantum pressure contribution proportionally to $k^4$ while usual pressure contributions modifies the evolution of the density contrast (which will be defined soon) according to $k^2$ \cite{Chavanis}. Therefore, the quantum pressure corrections could be relevant for the very small scales. Besides, in the top-hat spherical collapse the density of all fluids inside the spherical overdense region is homogeneous \cite{Abramo2} and the anisotropic pressure (\ref{anisotropic}) should be zero.

A cosmological dark matter fluid with the above pressure leads to the background expansion 
\begin{equation}
H^2=\frac{8\pi G}{3}\left(\rho_b+\rho_{\chi}+\rho_{\Lambda}\right).
\end{equation}
where $\rho_{\chi}$ is the BEC dark matter density, which in the Thomas-Fermi limit is determined by the pressure (\ref{Pbec}) via the continuity equation.

More details will be discussed in sections \ref{SectionIV} and \ref{SectionV}. In fact, we will follow in this work the background expansion determined in ref. \cite{harko1}.

\section{The nonlinear top-hat collapse}\label{SectionIII}

Here we present the basic equations that describe the evolution of a spherical collapsing matter region in an expanding background. This is the ideal technique for studying the clustering patterns of dark matter halos.

We will follow standard calculations presented in Refs. \cite{Abramo:2007iu, Abramo2, Rui, carames}. For general fluids, we define quantities such as
\begin{eqnarray}
\vec{v}^c &=& \vec{u}^0 + \vec{v}^p, \\
\rho^c&=&\rho\left(1+\delta\right) , \\
p^c&=&p + \delta p.
\end{eqnarray}
They are respectively the velocity, density and pressure of the collapsed region. The background velocity expansion is given by $\vec{u}^0$ and is associated with the Hubble's law. Peculiar motions are denoted by $\vec{v}^p$. The total density within this spherical region under collapse $\rho^c$ is written as the sum of the background density and the overdensity fraction $\delta \rho$. The same happens to the pressure definition.

The rate at which the overdense region expands reads
\begin{equation}
h=H+\frac{\theta}{3}(1+z),
\end{equation}
where $\theta=\vec{\nabla} \cdot \vec{v}^p$.

Energy conservation is also required for the collapsing region. Therefore, each component $i$ obeys a separate equation of the type
\begin{equation}\label{eqdeltaorig}
\dot{\delta_i}=-3H(c^2_{eff_i}-w_i)\delta_i-\left[1+w_i+(1+c^2_{eff_i})\delta_i\right]\frac{\theta}{a},
\end{equation}
where the energy density contrast is defined as
\begin{equation}\label{deltadef}
\delta_i = \left(\frac{\delta \rho}{\rho}\right)_i,
\end{equation}
and the effective speed of sound is computed following $c^2_{eff_i} = (\delta p / \delta \rho)_i$. Note that over dot means derivative with respect to the cosmic time $\it t$.

The dynamical evolution of the homogeneous spherical region will be governed by the Raychaudhuri equation
\begin{equation}\label{eqthetaorig}
\dot{\theta}+H\theta+\frac{\theta^2}{3a}=-4\pi Ga \sum_i (\delta\rho_i + 3\delta p_i)\ .
\end{equation}

For a cosmological model composed by $N$ distinct fluids one has to solve $N+1$ equations. One of the type \ref{eqdeltaorig} for each fluid and, since we adopt the top-hat profile, one single equation for the velocity potential $\theta$ which is sourced by the density fluctuations of the $N$ fluids.

Since we will use the  standard $\Lambda$CDM universe as our reference model here we show its equations for the spherical collapse. Both the baryonic and the dark matter component are assumed to be pressureless fluids. Therefore, we can write down

\begin{equation} \label{blambda}
\dot{\delta}_b=-\left(1+\delta_b\right)\frac{\theta}{a}, 
\end{equation}
\begin{equation} \label{dmlambda}
\dot{\delta}_{dm}=-\left(1+\delta_{dm}\right)\left(1+\sigma^2\right)\frac{\theta}{a},
\end{equation}
\begin{equation} \label{thetalambda}
\dot{\theta}+H\theta+\frac{\theta^2}{3a}=-4\pi G a \left[\rho_b\delta_b+\rho_{dm}\delta_{dm}(1+\sigma^2)\right]\ .
\end{equation}
Note that there is no equation of clustering of the cosmological constant since it is treated as a background quantity. Therefore, it influences this set of equations only via the expansion rate $H\equiv H(\rho_b, \rho_{dm}, \Lambda)$. In order to numerically solve (\ref{blambda}-\ref{thetalambda}) one usually specifies the initial conditions for $\delta_b$, $\delta_{dm}$ and $\theta$ at the redshift of decoupling $z_{dec}\sim 1000$ from which one can treat baryons as an independent fluid.

\section{Abrupt phase transition} \label{SectionIV}

The temperature $T_{crt}$ sets the beginning of the BEC phase transition. This is in fact a process which takes some finite time $\Delta t$ until all the normal dark matter has been converted into the BEC phase. As estimated in \cite{harko1} $\Delta t$ is of order of $10^{6}$ years. Although the latter value is parameter dependent, it is in general indeed an almost negligible fraction of the universe's lifetime. Therefore, the assumption that at $z_{crt}$ there is an instantaneous conversion to the BEC phase seems to be plausible and it will be considered in this section.

For $z>z_{crt}$ the dark matter equation of state calculated in (\ref{pdm}) reads
\begin{equation}\label{p1}
   p_{dm} = \sigma^2\rho_{dm} \,,
\end{equation}
where $\sigma^2 \equiv \left\langle \vec{v}^2 \right\rangle / 3c^2$. Applying this to the continuity equation one finds
\begin{equation}\label{rho1}
   \rho_{\chi} = \rho_{crt}\left(\frac{1+z}{1+z_{crt}}\right)^{3(1+\sigma^2)} \,, \quad z\geq z_{crt} \, ,
\end{equation}
where $z_{crt}$ is the redshift at the transition point and $\rho_{crt}\equiv\rho(z_{crt})$.

 For $z<z_{crt}$ the effective equation of state of the BEC dark matter is
\begin{equation}
   p_\chi = u_0 \rho_\chi^2 \,, \quad u_0 \equiv \frac{2 \pi \hbar^2 l_s}{m^3_{\chi}} \, ,
\end{equation}
and again, using the continuity equation we find
\begin{equation}
   \rho_\chi = \frac{\rho_{crt}}{(1+\omega_{crt})(\frac{1+z_{crt}}{1+z})^3-\omega_{crt}} \, , \quad z\leq z_{crt} \,,
\end{equation}
where $\rho_\chi$ is a continuous function at $z_{crt}$ and $\omega_{crt} \equiv \frac{p_{crt}}{\rho_{crt}}=\sigma^2$. At this point the continuity of the pressure (see discussion in \cite{harko1}) sets
\begin{equation}
   \label{eq:pressaoconstante}
   \sigma^2 \rho_{crt} = u_0 \rho_{crt}^2 \quad \Rightarrow \quad \rho_{crt} = \frac{\sigma^2}{u_0} \, ,
\end{equation} 
which, of course, depends on the model parameters. From this definition,
\begin{equation}
   \rho_{\chi 0} = \frac{\rho_{crt}}{(1+\omega_{crt})(1+z_{crt})^3-\omega_{crt}} 
\end{equation}
resulting in
\begin{equation}\label{redshiftcrt}	
	(1+z_{crt})^3 = \frac{\frac{\Omega_{crt}}{\Omega_{\chi 0}}+\omega_{crt}}{1+\omega_{crt}} \, ,
\end{equation}
where $\Omega_{\chi 0} = 0.25$ is the today's fractionary dark matter energy density parameter. The critical temperature at the point of Bose-Einstein condensation is
\begin{eqnarray}
   &T_{crt}&=\frac{2\pi \hbar^2 \rho_{crt}^{2/3}}{\zeta(3/2)^{2/3}m_\chi^{5/3}k_B} =\left[\frac{2\pi \hbar^2}{\zeta(3/2)^2 k_B^3}\frac{m_\chi\sigma^4}{l_s^2}\right]^{1/3} \\
	        &=& 6.87 \left(\frac{m_\chi}{1~\textup{meV}}\right)^{1/3}\left(\frac{\sigma^2}{3\times 10^{-6}}\right)^{2/3}\left(\frac{l_s}{1~\textup{ftm}}\right)^{-2/3}\textup{eV} \,,
					\nonumber
\end{eqnarray}
where $\zeta(3/2)$ is the Riemann zeta function and $k_B$ is the Boltzmann constant.

Note that before the phase transition we have $c_s^2=\sigma^2 = \omega_{crt}$. After this point, the equation of state parameter and the adiabatic ($c^2_s=\partial p / \partial \rho$) speed of sound associated to this fluid reads, respectively,
\begin{equation}
   \omega_\chi(z) = u_0 \rho_\chi(z) \, , \quad c^2_{s_\chi} = 2u_0 \rho_\chi(z) = 2\omega_\chi(z) \, .
\end{equation}

Concerning the perturbed region the effective speed of sound is actually given by the expression
\begin{equation}\label{ceffchi}
c^2_{eff_{\chi}}=\frac{\delta p_\chi}{\delta \rho_\chi}=\frac{p^c_\chi-p_\chi}{\rho^c_\chi-\rho_\chi}=w_\chi\frac{(1+\delta_\chi)^2-1}{\delta_\chi}=w_\chi(2+\delta_{\chi}),
\end{equation}
from which one can expand for small values of $\delta$ finding $c^2_{eff} \rightarrow c^2_s$ as expected.

\begin{figure}
\begin{center}
\includegraphics[width=0.47\textwidth]{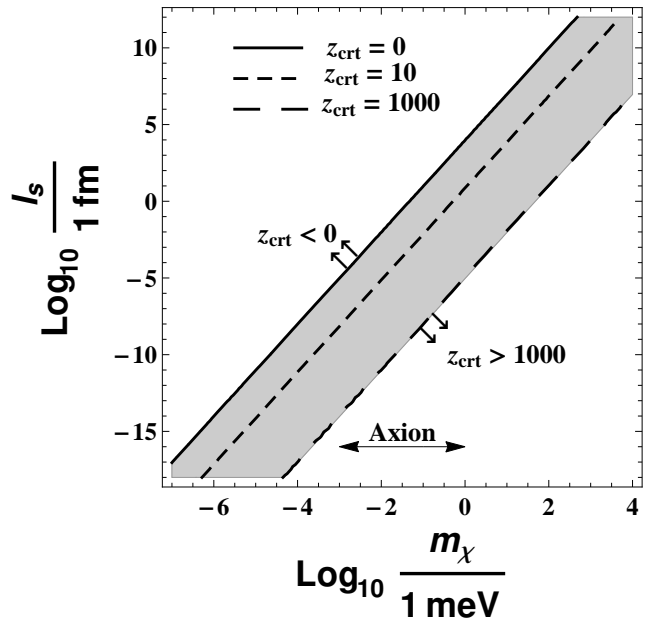}
\caption{The redshift of the phase transition ($z_{crt}$) in the parameters plane $l_s$ x $m$. The solid line sets the parameters in which the transition happens today at $z=0$. The long dashed line sets the parameters for which the transition takes place around the decoupling time $z_{cr}=1000$. The axion mass range is shown only for the sake of comparison.}
\label{zcA}
\end{center}
\end{figure}

Since a crucial issue in this model is the determination of the moment at which the transition happens in Fig. \ref{zcA} we show the dependence of $z_{crt}$ on the model parameters $m_{\chi}$ and $l_s$. This figure is numerically done after solving the equality proposed in (\ref{redshiftcrt}). A giving $z_{crt}$ value represents a curve in the $m_{\chi}$ {\it versus} $l_s$ plane. The solid line sets the parameter values for which the transition happens today at $z=0$. Therefore, only for the parameters values below the solid line the BEC dark matter model is able to leave some imprint on the observations. Note, for example that the configuration $(m_{\chi} , l_s)=(10^{-4} $ meV $ , 10^5 $ fm$)$ is an acceptable one. However, is this case, it would be impossible to probe the bosonic nature of dark matter since the transition will happen in a far future. On the other hand, over the long-dashed line the transition happens at the time of photon-baryon decoupling. In principle, $z_{crt}<1000$ is also allowed but its possible effect on the primordial CMB anisotropies is still not clearly known. Although this issue has not yet been investigated in detail we keep for convenience $0<z_{crt}<1000$ where we can consider a matter dominated universe--apart from late $\Lambda$ effects--and pressureless baryons. This redshift range corresponds to the gray region in this plot. The short-dashed line corresponds $z_{crt}=10$ and it is shown to guide the reader on how $z_{crt}$ evolves in this plane.
 
We also show in this figure the usual range for axion masses $10^{-3}$meV  $ < m_{\rm axion} < 1 $meV. Taking typical axion scattering lengths $<10^{-16}$ fm, Fig. 1 estimates correctly that the axion condensation happens indeed very early in the universe history. In our work we are not advocating in favor of any specific DM particle candidate. But in particular it is desired that most of the successes of the standard CDM paradigm should be kept. Indeed, it has been realised long ago that axions are very promising candidates for CDM \cite{axionCDM}.
Therefore, the existence of such particles exemplifies the validity of our approach since it guarantees the non-relativistic behavior of the DM component before the phase transition takes place. Of course, there is no direct relation to actual CDM axion models, which condensate much earlier in the universe history, to our approach. Notice also that axions are characterized by an attractive self-interaction. We just use them as instance of CDM light particles. At the same time, our approach also relies on the fact that before the transition we are dealing with CDM like particles. Therefore, one should avoid to keep in mind the use of lighter particles since they would be associated to warm/hot dark matter models. 
 
The meaning of the mass of the dark matter particle is quite clear. But in the cosmological context what does the scattering length $l_s$ mean?

Typical BEC experiments work with values in the range $10^{6} $ fm$<l_s <10^{9}$ fm. For these values the condition $z_{crt}>0$, i.e., assuring that the transition has already occurred, is satisfied for masses $m > 10$ meV and $100$ meV, respectively. Of course, usual BEC experiments with atoms cannot guide us in our search for viable DM parameters. However, we also note that by extrapolating the contours to lighter particles, as for example ultra-light masses of order $m \sim 10^{-22}$ meV, $z_{crt}>0$ requires almost negligible $l_s$ values which can be much smaller than the Planck length ($l_{plk}\sim 10^{-20}$ fm).

It is also worth noting that the space parameter indicated by the gray region is consistent with the stability of BEC dark matter halos as calculated in Ref. \cite{haloenergy}. However, see also a related discussion on the non-stability of BEC halos in Ref. \cite{halo}.

In order to solve for the evolution of the perturbed quantities during the collapse we adopt the following strategy. We solve numerically the $\Lambda$CDM equations taking initial conditions at a redshift $z_i=1000$ and with the values $\delta_{dm}(z_i)=3.5 \times 10^{-3}$, $\delta_{b}(z_i)=10^{-5}$ and $\theta(z_i)=0$ \cite{Rui, carames}. These values represent the standard amplitudes in the linear perturbation spectrum associated to today's clusters scales around the decoupling time. Indeed, the top-hat profile remains appropriate for such scales. Notice that clusters scales collapsed at low redshifts and therefore already within the BEC dark matter epoch. Smaller scales which have collapsed before the BEC phase transition will preserve the CDM structure and only differences in the final virial configuration would exist which is not the scope of this work. With such initial conditions this set of equations is evolved until the critical redshift $z_{crt}$. At this point, the quantities $\delta_{dm}(z_{crt})$, $\delta_{b}(z_{crt})$ and $\theta(z_{crt})$ are used as initial conditions for the BEC dark matter equations, which uses \ref{ceffchi}, from the critical redshift to $z=0$.

\begin{figure}
\begin{center}
\includegraphics[width=0.43\textwidth]{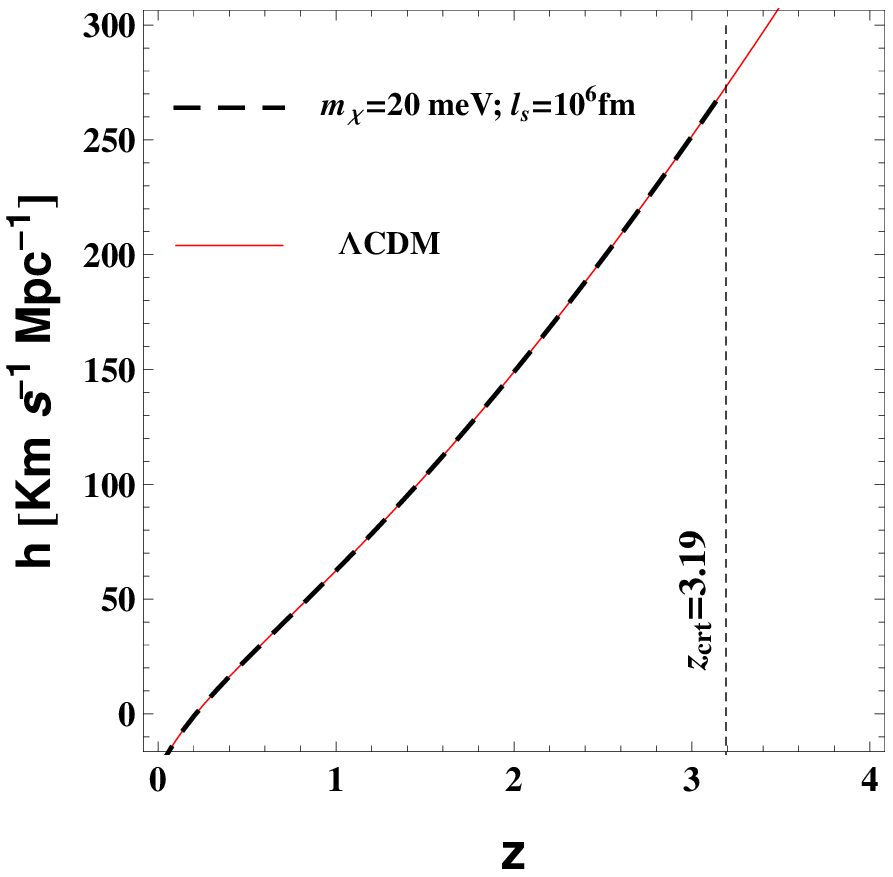}
\includegraphics[width=0.43\textwidth]{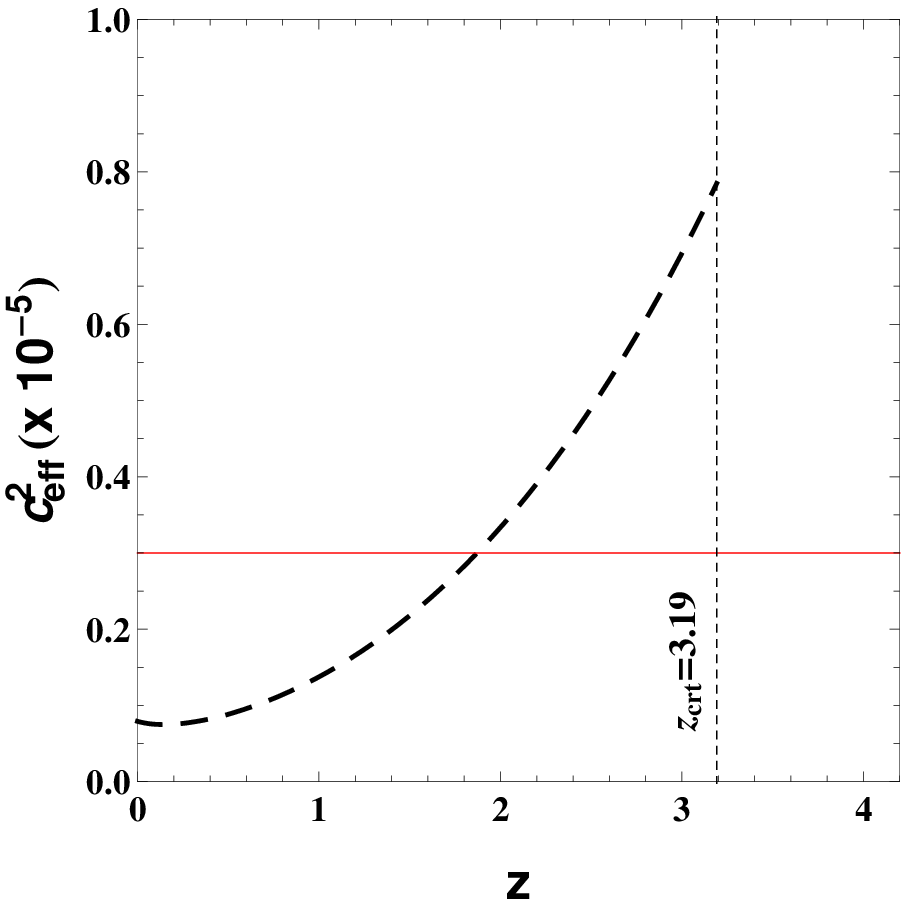}
\caption{Expansion rate (left) and effective speed of sound (right) of the collapsed region. In both plots we have $m_{\chi}=20$ meV and $l_s=10^6$ fm.}
\label{fig2}
\end{center}
\end{figure}

We have studied in great detail the parameter space $m_\chi$ and $l_s$ and although the BEC dark matter model indeed yields to a distinct dynamics at nonlinear level, this difference is, in practice, almost negligible. We show in the left panel of Fig. \ref{fig2} this feature where the expansion of the collapsed region is shown. The solid red line represents the standard cosmology while the dashed black line was calculated for a mass $m_{\chi}=20$ meV and a scattering length $l_s=10^6$ fm. With this choice the transition occurs at $z_{crt}=3.19$ as seen in the vertical dashed line. Both curves are in practice indistinguishable. The effective speed of sound is plotted in the right panel of Fig. \ref{fig2}. This shows the reason there are no significant changes in the evolution. We remark again that this result is not due to the specific choice $m_{\chi}=20$ meV and $l_s=10^6$ fm. It is a general feature of the model.

\section{Smooth phase transition}\label{SectionV}

We deal now with the situation in which there is a gradual conversion of ``normal'' dark matter into the condensed phase which starts at a redshift $z_{crt}$ and is finished at a redshift $z_{BEC}$. This is indeed the more realistic case. The dynamics shown in this section was also developed for the first time in Ref. \cite{harko1}.

\begin{figure}
\begin{center}
\includegraphics[width=0.43\textwidth]{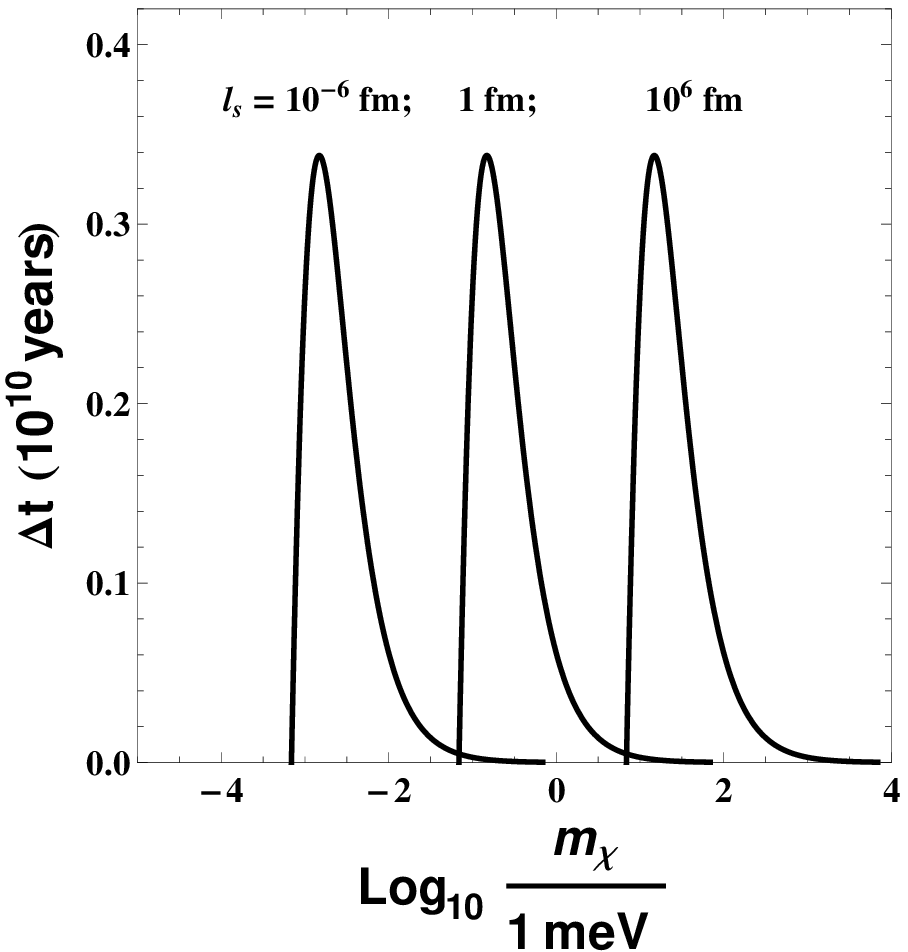}
\includegraphics[width=0.43\textwidth]{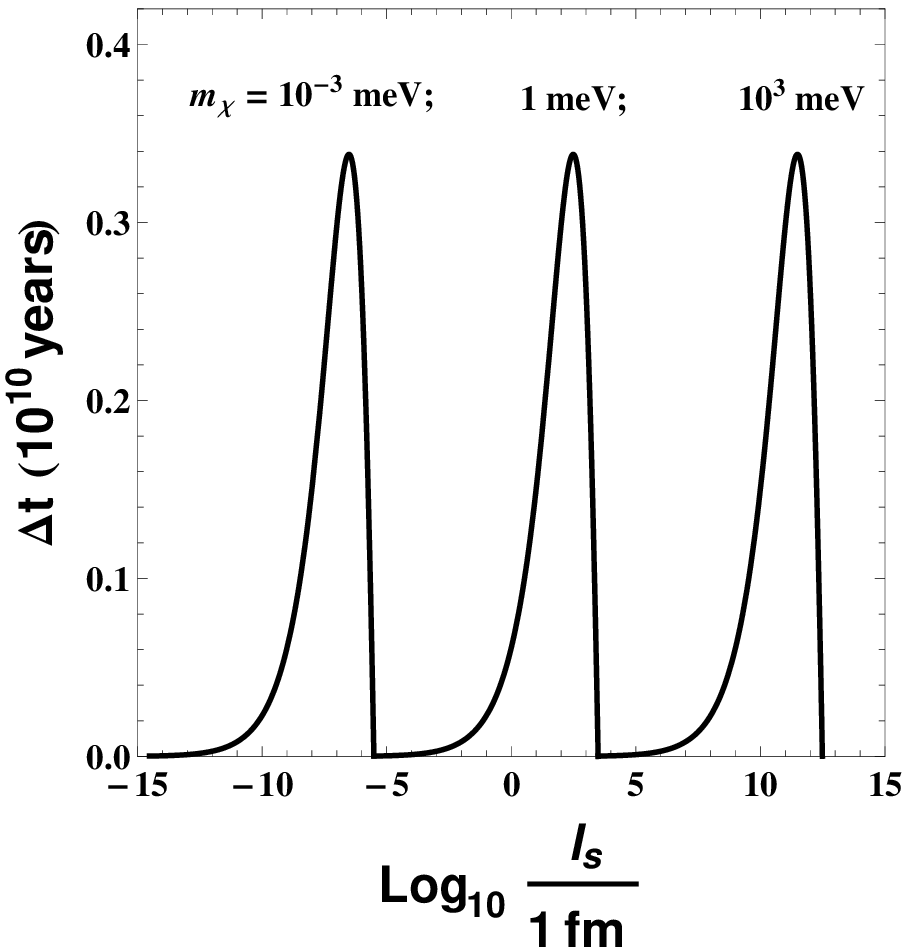}
\caption{The phase transition time length $\Delta t$ as a function of the models parameters $l_{\mathrm s}$ and $m$, where we fixed $\sigma^2=3 \times 10^{-6}$. }
\label{fig.Deltat}
\end{center}
\end{figure}

As mentioned in the last section the estimated duration $\Delta t=t(z_{BEC})-t(z_{crt})$ of this transition is of order $\Delta t \sim 10^6$ years, that is a small fraction of the universe's lifetime $t_U\sim 10^{10}$ years \cite{harko1}. However, $\Delta t$ depends on the model parameters $l_s$ and $m_{\chi}$. We calculate here again $\Delta t$ for some values $l_s$ and $m_{\chi}$ and plot it in Fig. \ref{fig.Deltat}. In the right panel of this figure, there is a maximum value $\Delta t_{\mathrm{max}} = 3.4\times 10^{9}$ years assuming, for instance, a mass $m_{\chi} = 1$ meV and $l_{s} \sim 3.1\times 10^2~\mathrm{fm}$. There are of course other combinations of $l_s$ and $m_{\chi}$ which produces similar $\Delta t$ values. The lower values for $\Delta t$ we have found are $\sim 10^6$ years. Therefore, this analysis shows that contrary to previous estimations, the phase transition can last a non-negligible fraction of the universe's lifetime. It is worth noting that recently Ref. \cite{harko2015} has pointed out the preferred values $m_{\chi}\sim 10^{-3}$ meV and $l_s \sim 10^{-7} ~\mathrm{fm}$ which according to our Fig. 3 maximize the duration of the phase transition.

As we will see below, the background dynamics and the evolution of the perturbation for the smooth phase transition differs significantly from the abrupt case studied in the last section. Then, one can expect that now we can observe some distinguishable feature of the BEC dark matter nonlinear collapse.

Let us now develop the dynamics during the smooth phase transition.
Before the transition starts, we have the same dynamics of a isotropic non-relativistic gas, as described in the last section by equations (\ref{p1}) and (\ref{rho1}). 
 
During the phase transition we can define the fraction of converted dark matter as
\begin{equation}
   \label{eq:deffracao}
   f(z) = \frac{\rho(z)-\rho_{crt}}{\rho_{\mathrm{BEC}}-\rho_{crt}} \,,
\end{equation}
where $\rho(z)$ is the dark matter density along the transition, $\rho_{crt}$ is the dark matter density before the transition and $\rho_{\mathrm{BEC}}$ its value afterwards. The function $f(z)$ is defined in such a way that at $z_{crt}$ we have $f(z_{crt})=0$. When the dark matter has fully converted to the BEC phase $f(z_{\mathrm{BEC}}) = 1$. 

Using (\ref{eq:deffracao}) into the continuity equation and integrating it from $z_{crt}$ to $z\geq z_{\mathrm{BEC}}$ we find
\begin{equation}
   \label{eq:fracao}
   f(z) = \frac{1+\omega_{crt}}{\frac{\Omega_{\mathrm{BEC}}}{\Omega_{crt}}-1}\left[\left(\frac{1+z}{1+z_{crt}}\right)^3-1\right] \, .
\end{equation}

Then, the dark matter density evolution becomes now

\begin{eqnarray}
   &&\rho_\chi=\rho_{crt}\left(\frac{1+z}{1+z_{crt}}\right)^{3(1+\sigma^2)} \,, \quad z \geq z_{crt} \, ; \\
	 \label{eq:densBECuni}
	 &&\rho_\chi=\rho_{crt}\left\{1+(1+\omega_{crt})\left[\left(\frac{1+z}{1+z_{crt}}\right)^3-1\right]\right\}, \\ \nonumber
	&&\quad z_{crt} \geq z \geq z_{\mathrm{BEC}} \, ;  \\
	 \label{eq:BECcondensada}
	 &&\rho_\chi=\rho_0 \frac{(1+z)^3}{(1+\omega_0)-\omega_0(1+z)^3} \,, \quad z \leq z_{\mathrm{BEC}} \, .
\end{eqnarray}

We still have to the determine the redshift $z_{\mathrm{BEC}}$ when the phase transition is over. With the condition $f(z_{\mathrm{BEC}})=1$ inserted in (\ref{eq:fracao}) we find
\begin{equation}
    \label{eq:eqOmegaBEC}
    \left[ \frac{\Omega_{\mathrm{BEC}}}{\Omega_{crt}}-1 + (1+\omega_{crt})\right]\left(\frac{1+z_{crt}}{1+z_{\mathrm{BEC}}}\right)^3 = (1+\omega_{crt}) \, ,
\end{equation}
and using $z=z_{\mathrm{BEC}}$ in the expression (\ref{eq:BECcondensada}) for the condensed dark matter density we have
\begin{equation}
    \label{eq:eqzBEC}		
	  \frac{\Omega_{\mathrm{BEC}}}{\Omega_{crt}} = \frac{\frac{\Omega_0}{\Omega_{crt}}(1+z_{\mathrm{BEC}})^3}{(1+\omega_0)-\omega_0(1+z_{\mathrm{BEC}})} \, .
\end{equation}
Eqs. (\ref{eq:eqOmegaBEC}) and (\ref{eq:eqzBEC}) can now be solved, leading to a solution for $z_{\mathrm{BEC}}$ and $\Omega_{\mathrm{BEC}}$.

As said before, during the phase transition both non-condensed and condensed dark matter coexist and the dark matter pressure is constant having the same value for both components in the interval $z_{\mathrm{BEC}}\leq z \leq z_{\mathrm{crt}}$, as given by Eq. \ref{eq:pressaoconstante}. We will assume that the same happens for the collapsed pressure $p^{\mathrm c}$. This allows us to find the constraint
\begin{equation}
   1+\delta_\sigma^{\mathrm{crt}} = (1+\delta_{\mathrm B}^{\mathrm{crt}})^2 \, ,
\end{equation}
where we used the expression $\rho_\chi(z) = \rho_\sigma(z)+\rho_B(z)$, which compared with Eq. (\ref{eq:fracao}) allows us to identify $\rho_\sigma(z) = \rho_{\mathrm{crt}}(1-f(z))$ as the non-condensed dark matter density and $\rho_B(z) = \rho_{\mathrm{BEC}}f(z)$ as density of the condensed state.

The continuity of dark matter fluid pressure enable us to treat both components as one single fluid also at perturbed level. In this case, the effective fluid sound velocity during the phase transition becomes

\begin{equation}
c^2_{eff_\chi} = \frac{p^c_{crt} - p_{crt}}{\rho_\chi \delta_\chi} = \sigma^2\frac{\rho_{\mathrm{crt}}}{\rho_\chi}\frac{\delta_{\mathrm{crt}}}{\delta_\chi} = \omega(z)\frac{\delta_{\mathrm{crt}}}{\delta_\chi} \,,
\end{equation}
where $\omega(z)=\omega_{\mathrm{crt}}\rho_{\mathrm{crt}}/\rho_\chi(z)$ is the equation of state parameter for the dark matter fluid during the phase transition. After the phase transition is completed, i.e., when $z \leq z_{\mathrm{BEC}}$, the effective fluid sound velocity of the BEC dark matter will be
\begin{equation}
    c^2_{eff_\chi} = \frac{p^c_{crt} - p_{crt}}{\rho_\chi \delta_\chi} = \frac{\sigma^2\rho_{crt}\delta_{crt}}{\rho_\chi \delta_\chi} = \omega(z)(2+\delta_\chi) \,,
\end{equation}
where $\omega(z)=\omega_{\mathrm{crt}}\rho_\chi(z)/\rho_{\mathrm{crt}}$ is the equation of state parameter for the dark matter after the phase transition.

Since the velocity dispersion $\sigma^2$ for the dark matter particles before the BEC phase transition is small the same assumptions made on $z_{\mathrm{crt}}$ in the previous section is still valid here, and we will consider values for the model parameters ($m_\chi,l_{\mathrm s}$) such that $0<z_{\mathrm{crt}}<1000$. We will also consider only cases where $z_{\mathrm{BEC}}\geq0$.

This set of equations is evolved until the critical redshift $z_{crt}$ assuming the same initial conditions at $z_i$ as before. At this point, the quantities $\delta_{dm}(z_{crt})$, $\delta_{b}(z_{crt})$ and $\theta(z_{crt})$ are used as initial conditions for the phase transition perturbed Eqs. (\ref{eqdeltaorig}) and (\ref{eqthetaorig}) with the suitable background parameters. This set of equations is again evolved until the $z_{\mathrm{BEC}}$, and the quantities $\delta_{dm}(z_{\mathrm{BEC}})$, $\delta_{b}(z_{\mathrm{BEC}})$ and $\theta(z_{\mathrm{BEC}})$ are used as initial conditions for the BEC dark matter perturbed equations.

\begin{figure}
\begin{center}
\includegraphics[width=0.43\textwidth]{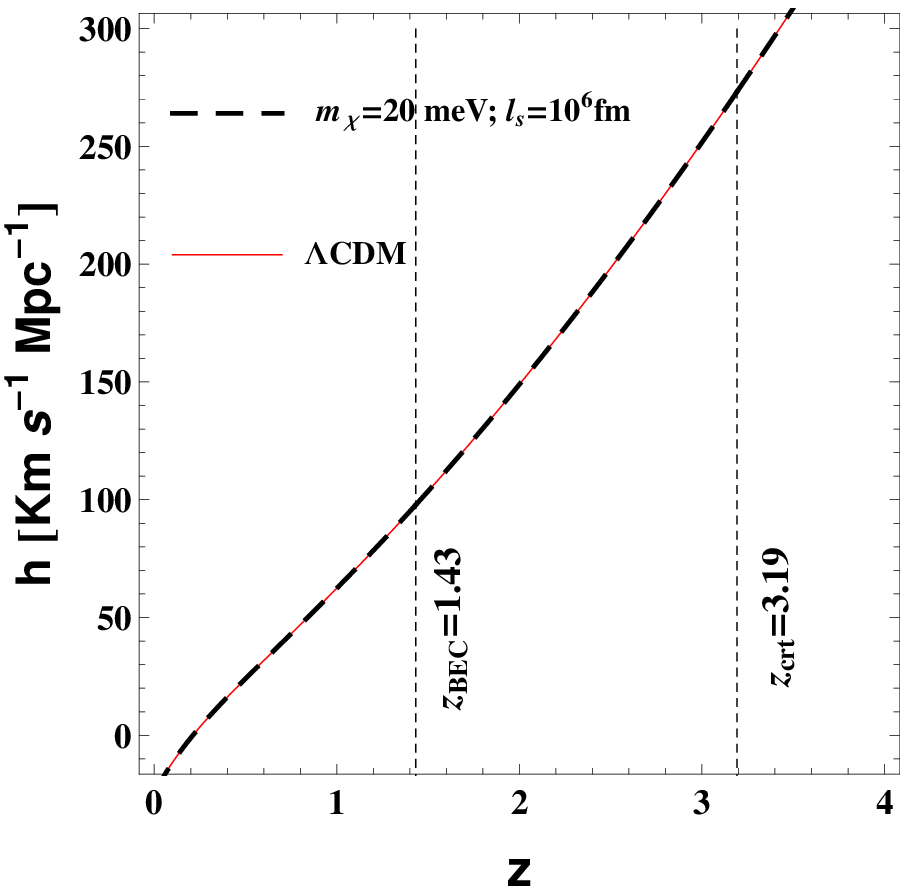}
\includegraphics[width=0.43\textwidth]{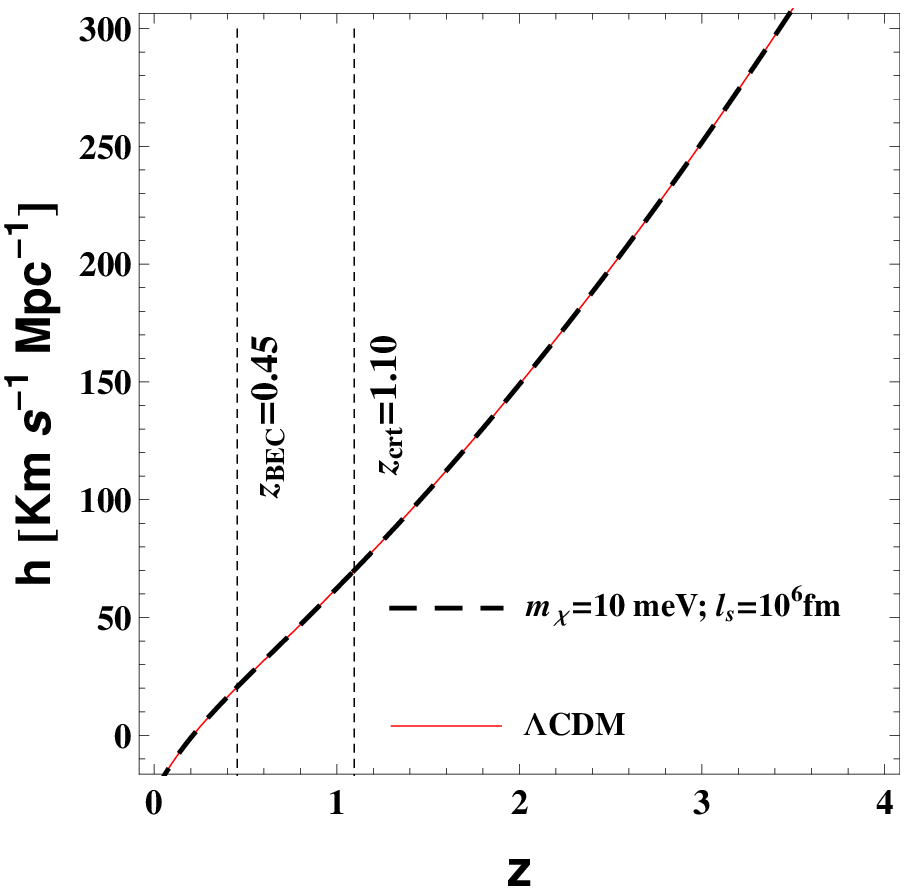}
\caption{Expansion rate of the collapsed region for the smooth phase transition approach.}
\label{fig.hSmooth}
\end{center}
\end{figure}

\begin{figure}
\begin{center}
\includegraphics[width=0.43\textwidth]{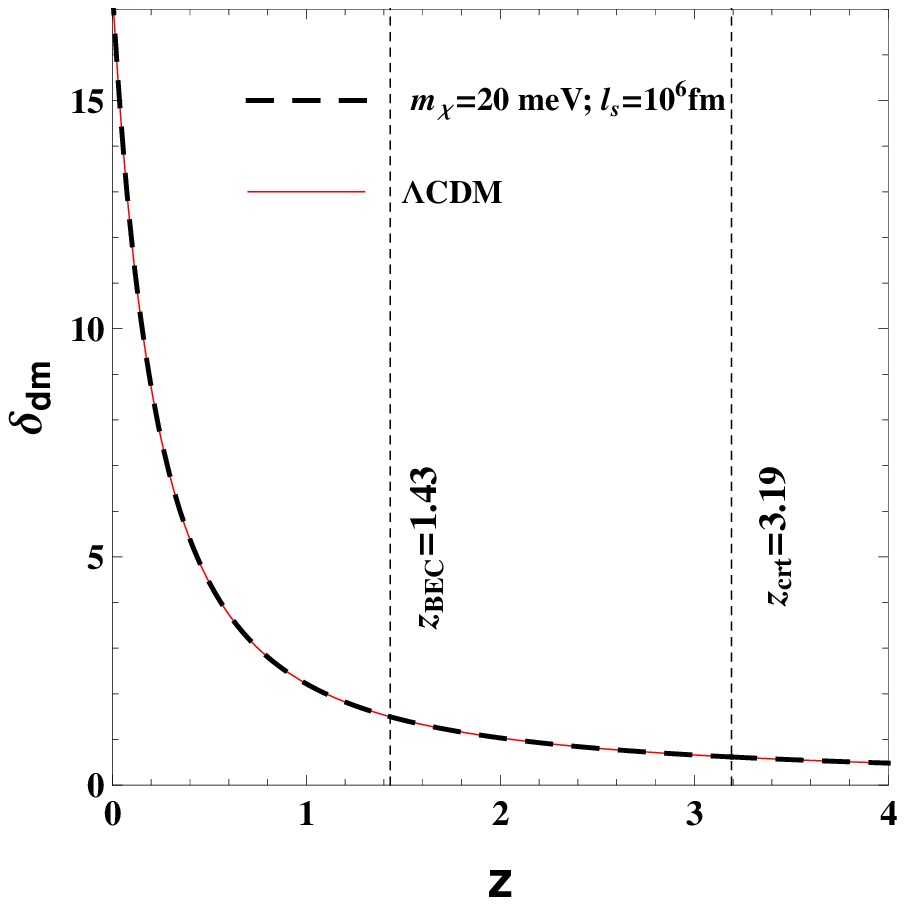}
\includegraphics[width=0.43\textwidth]{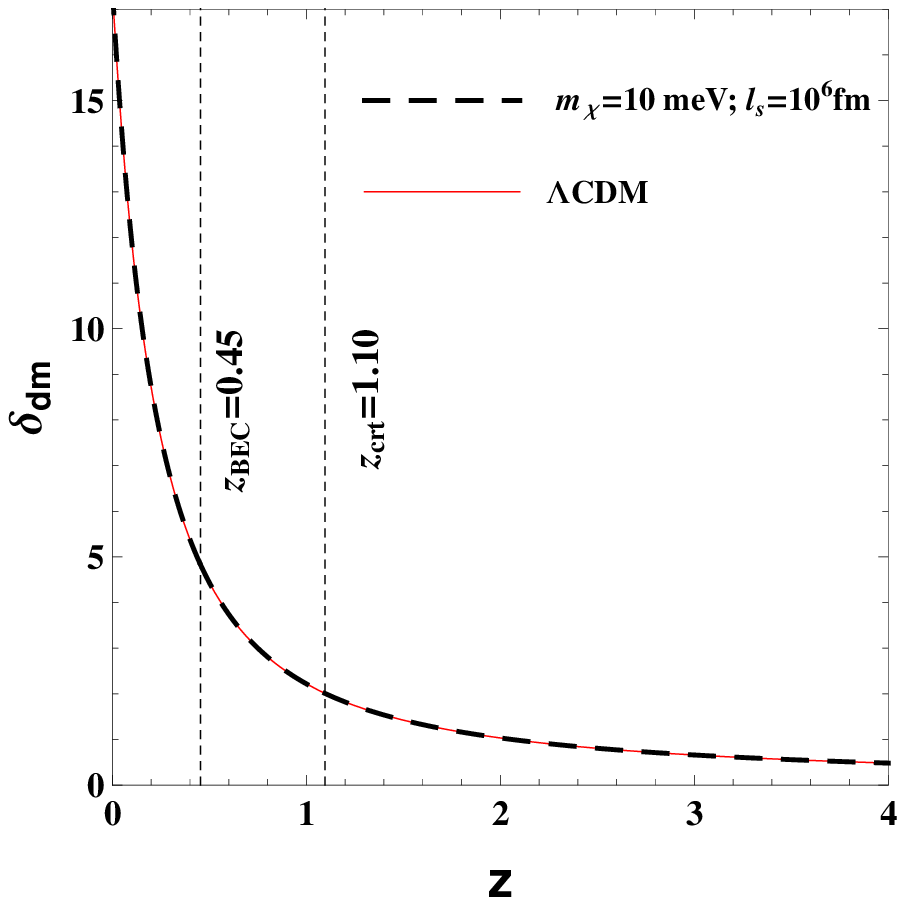}
\caption{Dark matter density contrast for the smooth phase transition approach.}
\label{contrast1}
\end{center}
\end{figure}

In Fig. \ref{fig.hSmooth} we show the expansion of the collapsed region for the smooth phase transition model, where the solid red curve represents the standard $\Lambda$CDM model and the black dashed curve represents the BEC model, for $m_\chi = 20$ meV in the left panel, $m_\chi=10$ meV in the right panel and $l_\mathrm{s} = 10^6$ fm in both cases. The dashed vertical lines show the initial and the final points of the transition phase. In the left panel, $z_\mathrm{crt}=3.19$ and $z_\mathrm{BEC}=1.43$ and $z_\mathrm{crt}=1.10$ and $z_\mathrm{BEC}=0.45$ in the right panel. These intervals correspond to $2.40\times 10^9$ years and $3.38\times 10^9$ years. As in the abrupt transition model there are no major difference between CDM and BEC dark matter.

The evolution of the non-linear density perturbations are shown in Fig. \ref{contrast1}, where $\delta_\mathrm{dm} \equiv \delta \rho_{\mathrm{dm}}/\rho_{\mathrm{dm}}$ is the dark matter density contrast. Again, the red curve represents the standard $\Lambda$CDM model, while the black dashed curve shows the behavior of the BEC model for $m_\chi = 20$ meV in the left panel, $m_\chi = 10$ meV in the right panel and $l_\mathrm{s} = 10^6$ fm in both cases. The curves are again indistinguishable. 

The redshift of turnaround $z_{\mathrm{ta}}$ is the one which marks the moment when the perturbed region starts to decrease its physical radius. This happens when $h=0$, i.e., $z_{\mathrm{ta}}=z(h=0)$. For $\Lambda$CDM model $z_{\mathrm{ta}}^{\Lambda\mathrm{CDM}} = 0.2113$ and for the cases seen in both panels of Fig. \ref{fig.hSmooth} we have $|z_{\mathrm{ta}}^{\mathrm{BEC}}-z_{\mathrm{ta}}^{\Lambda \mathrm{CDM}}| \approx 10^{-4}$.

\section{Conclusions}

We have studied the nonlinear clustering properties of the Bose-Einstein dark matter model. In this scenario, bosonic dark matter particles are able to undergo a phase transition as their temperature reaches the critical one $T_{crt}$ which corresponds to some critical redshift $z_{crt}$. The main questions here are: i) how does $z_{crt}$ depend on the fundamental model parameters $m_{\chi}$ (the particle mass) and $l_s$ (the scattering length)? and ii) what is the background and perturbative dynamics during the phase transition?

Fig. \ref{zcA} shows in detail the expected degeneracy of $z_{crt}$ values in the $l_s$ x $m_{\chi}$ plane, i.e., for a given $z_{crt}$, there are many admissible parameter configurations. This result identifies the parameters values for which $z_{crt}>0$ and therefore are able to leave imprints on large scale structure observations. At the same time, if the actual parameters values of the BEC model lie in the region $z_{crt}<0$ then the bosonic nature of the dark matter particles cannot be accessed via cosmological observables. If the present model is employed for BEC phase-transitions, ultra-light candidates ($m_{\chi} \lesssim 10^{-22}$ eV) would only lead to possible observational imprints for $l_s$ of order of the Planck length or smaller. Note that this claim is limited to the fluid description used here. Recent calculations on the full dynamics of the ultra-light axion scalar field sho that there are indeed possible observable imprints in the cosmological data \cite{Pedro}.

Our strategy was to identify specific signatures of the BEC dark matter nonlinear clustering. Since there is a positive pressure associated to the BEC dark fluid one can expect that the corresponding effective speed of sound will modifies somehow the agglomeration rate. We tried to understand this process via both the abrupt and the smooth phase transition approaches. In the former scenario the dark matter dynamics changes suddenly at $z_{crt}$. In the latter, there is a continuous conversion from the ``normal'' to the BEC phase. Although we showed that the smooth transition can indeed last quite a significant fraction of the universe lifetime. Then, it seems that this case could, eventually, lead to a remarkable dynamics. However, in both approaches of the phase transition we could not identify any relevant difference between the BEC model and the standard CDM model. This is mostly because the model parameters leading to $z_{crt}<0$ produce almost negligible $c_{eff}$ values. On one hand, this guarantees that the nonlinear clustering patterns of the BEC model at large scales are very similar to the CDM model. We have provided a theoretical confirmation for the recent numerical results of Schive et al \cite{Schive} which claims that the differences between BEC DM and standard CDM appears only in the internal structure of DM halos rather than on the cosmological large scale distribution. On the other hand, this eliminates the cosmological nonlinear perturbative study as a possible technique to probe the bosonic nature of dark matter particles. It is also worth noting that the typical value for the critical overdensity for collapse $\delta_c=1.686$ remains unchanged for the BEC parameter space probed here. Perharps, this conclusion is in part due to the fact we have assumed a simple version of the first order phase transition of the BEC DM model. Taking properly into account, for example, the latent heat released during the transition and the resulting dynamics associated to the nucleation of the new bubbles we could end up with a very drastic effect on the non-linear clustering. We will leave this analysis for a future work.

{\bf Acknowledgments:}
We acknowledge T. Harko for useful correspondence and the anonymous referee for his/her remarks that substantially improved this work. We thank CNPq (Brazil) and FAPES (Brazil) for partial financial support. HV also acknowledges the financial support of A*MIDEX project (n° ANR-11-IDEX-0001-02) funded by the ``Investissements d'avenir" French Government program, managed by the French National Research Agency (ANR).


\begin{thebibliography}{00}

\bibitem{PlanckCosmoParam} P. A. R. Ade, {\it et al.}, arXiv:1502.01589 [astro-ph.CO].

\bibitem{Baer} H. Baer, K. Y. Choi, J. E. Kim and L. Roszkowski, Phys. Reports, {\bf 555}, 1 (2015).

\bibitem{cdmpressure} L. Xu, Y. Chang, Phys. Rev. D {\bf 88}, 127301 (2013); R. Hlozek, D. Grin, D. J. E. Marsh and P. G. Ferreira  Phys. Rev. D
{91}, 103512 (2015).

\bibitem{NFW} J. F. Navarro, C. Frenk, S. White, The Astrophysical Journal {\bf463}, 563 (1996).

\bibitem{gov} D. H. Weinberg, J. S. Bullock, F. Governato, R. Kuzio de Naray, A. H. G. Peter, arXiv:1306.0913v1 [astro-ph.CO]; J. Oñorbe {\it et. al.} arXiv:1502.02036v1 [astro-ph.GA].

\bibitem{Mod} A. De Felice and S. Tsujikawa, Living Rev. Relativity {\bf13}, 3 (2010); S. Capozziello, M. De Laurentis Physics Reports, {\bf 509} 4, 167 (2011); T. Clifton, P. G. Ferreira, A. Padilla, C. Skordis, Physics Reports {\bf513}, 1 (2012). 

\bibitem{Will} C.~M.~Will,
  Living Rev.\ Rel.\  {\bf 17}, 4 (2014).

\bibitem{HDM} S. Tremaine and J. E. Gunn. Phys. Rev. Lett. {\bf42}, 407 (1979).
 
\bibitem{wdm} Paul Bode {\it et al.} ApJ {\bf556}, 93 (2001); H. J. de Vega, P. Salucci, N. G. Sanchez, New Astronomy, {\bf17}, 653 (2012); C. Destri, H. J. de Vega, N. G. Sanchez, Phys. Rev. D{\bf88}, 083512 (2013); M. Viel, G.D. Becker, J.S. Bolton and M.G. Haehnelt, Phys. Rev, D {\bf 88} 043502 (2013).

\bibitem{wdm2} A. Schneider, D. Anderhalden, A. Maccio, J. Diemand, Mon.Not.Roy.Astron.Soc. {\bf441}, 6 (2014).

\bibitem{fuzzy}  W. Hu, R. Barkana and A. Gruzinov. Phys. Rev. Lett. {\bf85}, 1158 (2000).

\bibitem{sidm} M. Rocha, A. H. G. Peter, J. S. Bullock, M. Kaplinghat, S. Garrison-Kimmel, J. Onorbe and L. A. Moustakas, MNRAS {\bf430}, 81 (2013). 

\bibitem{vdm} H. Velten, D. J. Schwarz, J. C. Fabris, W. Zimdahl, Physical Review D {\bf88}, 103522 (2013); H. Velten , IJGMMP {\bf11}, 02, 1460013 (2014); H. Velten, T.R.P. Caram\^es, J. C. Fabris, L. Casarini, R. C. Batista, Phys. Rev. D {\bf90}, 123526 (2014). 

\bibitem{BEClab}  C. C. Bradley, C. A. Sackett, J. J. Tollett, and R. G.
Hulet, Physical Review Letters
{\bf75}, 1687 (1995); M. H. Anderson, J. R. Ensher, M. R. Matthews, C. E.
Wieman, and E. A. Cornell, Science {\bf269}, 198 (1995); 

\bibitem{BEClab2} E. A. Cornell and C. E.
Wieman, Rev. Mod. Phys. {\bf74}, 875 (2002); W. Ketterle,
Rev. Mod. Phys.{\bf74}, 1131 (2002).

\bibitem{Siki} P. Sikivie and Q. Yang, Phys.Rev.Lett. {\bf103}, 111301 (2009).

\bibitem{Bohmer} C.G. B\"ohmer and T. Harko, JCAP {\bf06}
(2007) 025.


\bibitem{harko1} T. Harko, Phys. Rev. D {\bf83}, 123515 (2011).

\bibitem{Abril} A. Suarez, Victor H. Robles, T. Matos, Astrophysics and Space Science Proceedings {\bf 38}, Chapter 9 (2013). 

\bibitem{Maxim} M.Yu.Khlopov, A.S.Sakharov and D.D.Sokoloff, Nucl.Phys. B (Proc. Suppl.) {\bf72},
105-109 (1999); I.G.Dymnikova, M.Yu.Khlopov, Mod. Phys.
Lett. A {\bf15} 2305 (2000).

\bibitem{LiShapiro} B. Li, T. Rindler-Daller, and Paul R. Shapiro, Phys. Rev. {\bf D89} (2014) 083536. 

\bibitem{Guth}  A. H. Guth, M. P. Hertzberg, C. Prescod-Weinstein [arXiv:1412.5930 [astro-ph.CO]].



\bibitem{BEC} F. Dalfovo, S. Giorgini, L.P. Pitaevskii and S. Stringari, Rev. Mod. Phys. {\bf71} (1999) 463.
 
\bibitem{MaximZel} M. Yu. Khlopov, B. E. Malomed and Ya. B. Zeldovich,  MNRAS {\bf 215}, 575 (1985).

\bibitem{AbrilSF} A. Suarez, T. Matos, Mon.Not.Roy.Astron.Soc. {\bf416}, 87 (2011). 

\bibitem{wamba} H. Velten and E. Wamba, Phys.
Lett. B {\bf709}, 1 (2012).

\bibitem{Chavanis} P.-H. Chavanis, A\&A {\bf537},A127 (2012).
 
\bibitem{BenKain} B. Kain and H. Y. Ling, Phys. Rev. D {\bf85}, 023527 (2012).

\bibitem{rodolfo} R. C. Freitas and S. V. B. Gon\c calves, JCAP {\bf04}, 049 (2013). 

\bibitem{Alcu} M. Alcubierre, A. de la Macorra, A. Diez-Tejedor, J. M. Torres, arxiv:1501.06918 [gr-qc].

\bibitem{simu} V. Springel, Astronomische Nachrichten, {\bf333}, Issue 5-6, 515 (2012).

\bibitem{Mocz} P. Mocz and S. Succi, [arXiv:1503.03869 [physics.comp-ph]].

\bibitem{Schive} H.-Y. Schive , T. Chiueh and T. Broadhurst, Nature Physics, {\bf10}, 496 (2014); H.-Y Schive, M.-H. Liao, T.-P Woo, S.-K. Wong, T.
Chiueh, T. Broadhurst and W.-Y.P. Hwang, Phys. Rev. Lett., {\bf113}, 261302 (2014).

\bibitem{HarkoBECCollapse} T. Harko, Phys. Rev. D {\bf89}, 084040, (2014).


\bibitem{Ju} P. S. Julienne, F. H. Mies, E. Tiesinga, and C. J.
Williams, Phys. Rev. Lett. {\bf70}, 1880 (1997).

\bibitem{Abdu}F. Kh. Abdullaev, B. B. Baizakov, S. A. Darmanyan, V.
V. Konotop, and M. Salerno, Phys. Rev. A {\bf 64}, 043606 (2001).

\bibitem{delfini} Pierre-Henri Chavanis, Phys. Rev. D {\bf 84}, 043531 (2011) ; P.H. Chavanis, L. Delfini, Phys. Rev. D {\bf84}, 043532 (2011). 

\bibitem{RindlerDaller} T.~Rindler-Daller and P.~R.~Shapiro, Mon.\ Not.\ Roy.\ Astron.\ Soc.\  {\bf 422}, 135 (2012).


\bibitem{Abramo2} L. R. Abramo, R. C. Batista, L. Liberato, and R. Rosenfeld, Phys. Rev. D {\bf 79}, 023516 (2009).



\bibitem{Abramo:2007iu}  L.~R.~Abramo, R.~C.~Batista, L.~Liberato and R.~Rosenfeld,
  JCAP {\bf 0711}, 012 (2007).
	
\bibitem{Rui} R. A. A. Fernandes, J. P. M. de Carvalho, A. Yu. Kamenshchik, U. Moschella, A. da Silva, Phys. Rev. D {\bf 85}, 083501 (2012).

\bibitem{carames} Thiago R. P. Caram\^es, J\'ulio C. Fabris, Hermano E. S. Velten, Phys. Rev. D {\bf89}, 083533 (2014); Hermano E. S. Velten, Thiago R. P. Caram\^es, Phys. Rev. D {\bf 90}, 063524 (2014).

\bibitem{axionCDM} J. Preskill, M. Wise, and F. Wilczek,
Phys. Lett. B {\bf120}, 127 (1983); L. Abbott and P. Sikivie,
Phys. Lett. B {\bf120}, 133 (1983); M. Dine and W. Fischler,
Phys. Lett. B {\bf120}, 137 (1983); P. Sikivie,
Lect. Notes Phys. {\bf741}, 19 (2008). 

\bibitem{haloenergy} J. C. C. de Souza, M. O. C. Pires, JCAP{\bf03},  010 (2014).	

\bibitem{halo} F. S. Guzman, F. D. Lora-Clavijo, J. J. Gonzalez-Aviles, F. J. Rivera-Paleo, JCAP {\bf09}, 034 (2013). 

\bibitem{harko2015} T. Harko, P. Liang, S.-D. Liang, G. Mocanu, {\it Testing the Bose-Einstein Condensate dark matter model at galactic cluster scale },  arXiv: 1510.06275.


\bibitem{Pedro} D. Marsch and P. Ferreira, Phys. Rev. D {\bf 82}, 103528 2010; R. Hlozek, D. Grin, D. J.E. Marsh, P. G. Ferreira, Phys. Rev. D {\bf 91}, 103512 (2015).


\end{thebibliography}
\end{document}